\documentclass[preprint,aps,amsmath,amssymb,floatfix]{revtex4-2}

\usepackage{graphicx}     
\usepackage{dcolumn}      
\usepackage{bm}           
\usepackage{hyperref}     
\usepackage{xcolor}
\graphicspath{{Figuras/}}
\usepackage{booktabs}
\graphicspath{{Figuras/}}

\begin{document}

\title{Statistical complexity from fluctuations 
in the information content}

\author{Renio S. Mendes}
\email{rsmendes@dfi.uem.br}
\affiliation{Departamento de F\'isica, Universidade Estadual de Maring\'a, Maring\'a, PR, Brazil}

\author{Sergio Picoli}
\email{spjunior@dfi.uem.br}
\affiliation{Departamento de F\'isica, Universidade Estadual de Maring\'a, Maring\'a, PR, Brazil}

\author{Evaldo M. F. Curado}
\email{evaldo@cbpf.br}
\affiliation{Centro Brasileiro de Pesquisas F\'\i sicas, Rio de Janeiro, RJ, Brazil}

\date{\today}

\begin{abstract}
We argue that the variance of the information content ($C$), an information-theoretic quantity, can be naturally interpreted as a measure of statistical complexity. We show that $C$ satisfies widely accepted criteria for statistical complexity measures: it vanishes for both ordered and equiprobable states, while attaining maxima in intermediate regimes, typically shifted toward order. This interpretation establishes direct connections with thermodynamics and phase transitions: for systems obeying Boltzmann--Gibbs statistics, $C$ is extensive and directly proportional to energy fluctuations and heat capacity. Moreover, unlike other statistical complexity measures, it attains a maximum at continuous phase transitions, as illustrated for the two-dimensional Ising model. Applications to chaotic maps and fractional Gaussian noise further indicate that $C$ captures nontrivial dynamical structure in different classes of correlated systems.
\end{abstract}

\maketitle

\section{Introduction}

The study of complex systems has extended the reach of physics to areas such as biology, economics, and social sciences, where emergent and multiscale behavior remains a central challenge~\cite{gao21}. In the characterization of such systems, entropy plays a central role as a measure of uncertainty or average information content~\cite{shannon48,renyi61,tsallis88,Khinchin53}. However, entropy itself does not capture the diversity of contributions from individual states: distinct probability distributions may share the same entropy while exhibiting different internal structures. 
This limitation has motivated the search for complementary descriptors of probabilistic structure and organization---although no universally accepted definition of complexity has emerged~\cite{lloyd01}. Some of these descriptors are based directly on the underlying probability distribution, including statistical complexity measures~\cite{crutchfield89,shepard92,landsberg99,calbet95,rosso04,ribeiro17}.

Suitable measures of statistical complexity are generally expected to vanish both in completely ordered states and at equiprobability, while attaining nonzero values in intermediate regimes~\cite{crutchfield89,landsberg99,feldman98}. 
Beyond these basic requirements, complexity maxima are often expected to occur closer to order than to complete randomness, reflecting the greater sensitivity of ordered states to perturbations. Another desirable property is extensivity, which has also been discussed in specific contexts~\cite{anteneodo96,landsberg99,feldman98,rosso04}.

In this work, we discuss a measure of statistical complexity $C$ defined as the variance of the information content---closely related to the Bates--Shepard proposal based on the standard deviation~\cite{shepard92}. The variance of the information content has been mainly explored in information theory and related contexts~\cite{verdu14,arikan16}, with recent applications in machine learning and time-varying systems~\cite{shternshis24,ahmed26}. Its role as a measure of statistical complexity, however, remains largely unexplored. We show that $C$ satisfies the general requirements expected for complexity measures and provides a simple and physically meaningful characterization of statistical complexity. For Boltzmann--Gibbs systems, $C$ is directly related to energy fluctuations and heat capacity and becomes extensive in the thermodynamic limit, establishing a connection between statistical complexity and thermodynamic response functions. Applications to the two-dimensional Ising model and chaotic maps further demonstrate that $C$ captures structural and dynamical features not fully described by entropy and representative statistical complexity measures.

\section{Definition and basic properties}

For a system with $W$ accessible states and probabilities ${p_\alpha}$, the Shannon entropy is $S = -\sum_\alpha p_\alpha\ln p_\alpha$. The complexity $C$ is then expressed as:
\begin{equation}\label{defC}
C = \langle(I - S)^2\rangle = \sum_\alpha p_\alpha(\ln p_\alpha)^2 - S^2,
\end{equation}
where $I_\alpha=-\ln p_\alpha$ is the information content (or surprise) associated with state $\alpha$, whose average corresponds to the Shannon entropy, $S=\langle I\rangle$. While entropy corresponds to the average information content, $C$ measures its variance, characterizing the heterogeneity of the probability distribution. Consequently, systems with identical entropy may exhibit different complexity values, motivating the use of the $S$-$C$ plane to distinguish structural and dynamical regimes.

\begin{figure}[t]
    \centering
    \includegraphics[width=0.49\columnwidth]{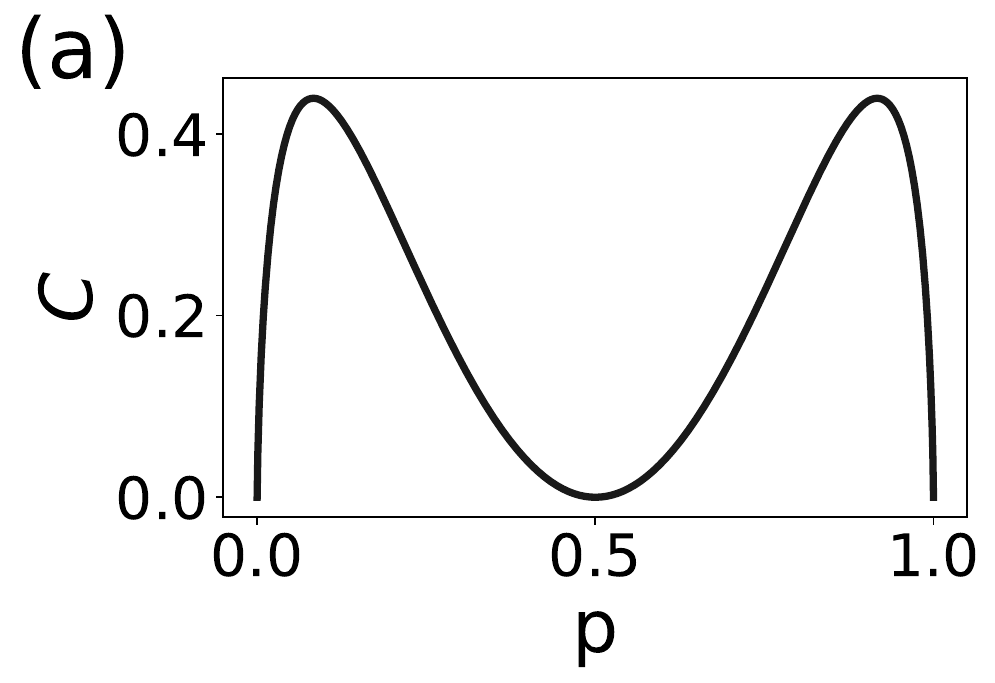}
    \hfill
    \includegraphics[width=0.49\columnwidth]{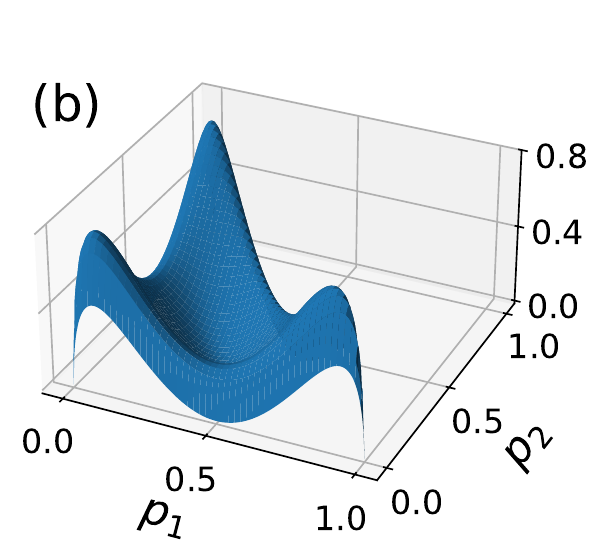}

    \vspace{2mm}

    \includegraphics[width=0.49\columnwidth]{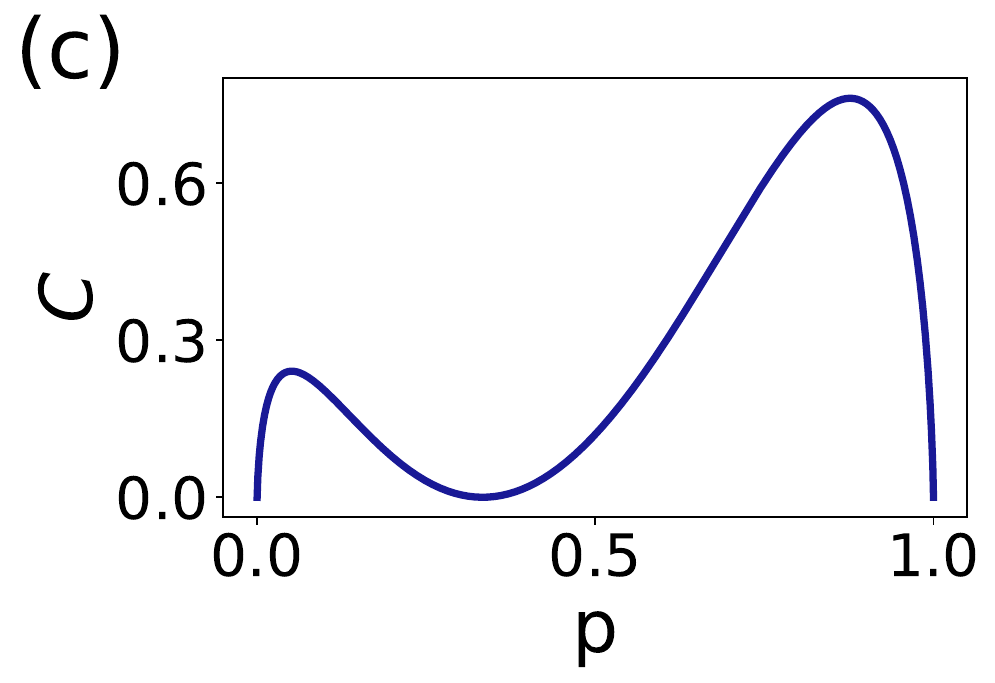}
    \hfill
    \includegraphics[width=0.49\columnwidth]{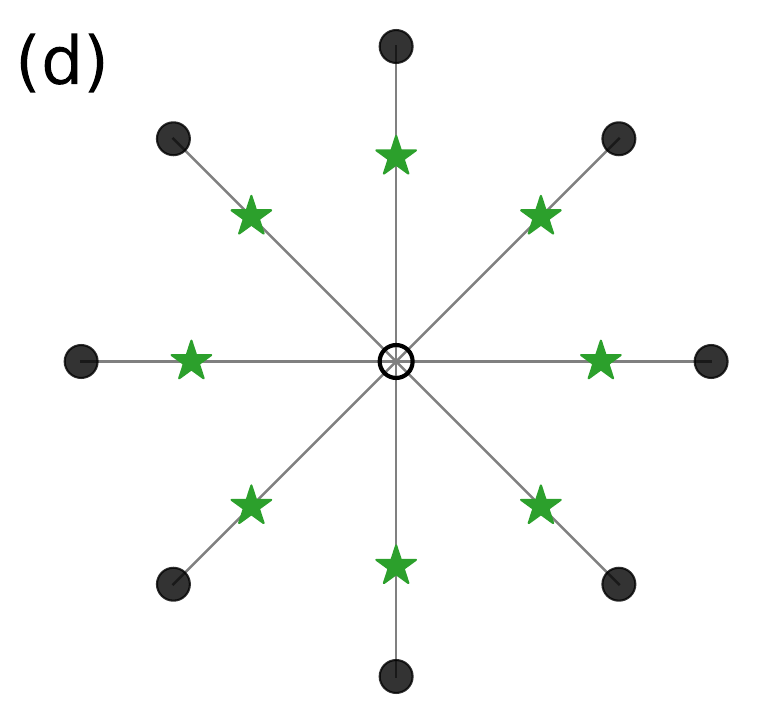}

    \caption{(a) $C(p)$ for two states. The maxima occur at $p^{*}\approx 0.9168$ and $1-p^{*}$, while $C=0$ at $p=1/2$. (b) $C(p_1,p_2)$ for three states. (c) Cut of the three-state simplex along $p_1=p$ and $p_2=(1-p)/2$, exhibiting a global maximum at $p^{*}\approx 0.8767$ and a secondary maximum near $p=0$. (d) Schematic representation for $W=8$. In all cases, complexity maxima (stars) occur closer to order (closed circles) than to equiprobability (open circle).}
    \label{fig1}
\end{figure}

To illustrate the basic properties of $C$, we examine first a two-state system with probabilities $p$ and $1-p$, for which Eq.~(\ref{defC}) becomes
\begin{equation}
C(p)=p(1-p)\left(\ln\frac{p}{1-p}\right)^2.
\end{equation}
As shown in Fig.~\ref{fig1}(a), $C(p)$ vanishes both in the ordered states ($p=0,1$) and at equiprobability ($p=1/2$), while exhibiting two symmetric maxima shifted toward order. Near $p=0$, $C \sim p(\ln p)^2$, implying a divergent derivative and a rapid departure from order, whereas near $p=1/2$ the behavior is smooth [$C \sim 4(p-1/2)^2$], indicating the stability of the equiprobable state. For three states, the same qualitative structure persists within the probability simplex ($p_1+p_2+p_3=1$): multiple maxima occur near ordered configurations, whereas a minimum appears at equiprobability [Fig.~\ref{fig1}(b)] (see also Ref.~\cite{paolillo21}). Along the symmetric path $p_1=p$ and $p_2=p_3=(1-p)/2$, the global maximum remains shifted toward order [Fig.~\ref{fig1}(c)], suggesting a general geometric organization in which complexity maxima lie between order and disorder but systematically closer to the ordered limit [Fig.~\ref{fig1}(d)].

\begin{figure}[t]
    \centering
    \includegraphics[width=0.49\columnwidth]{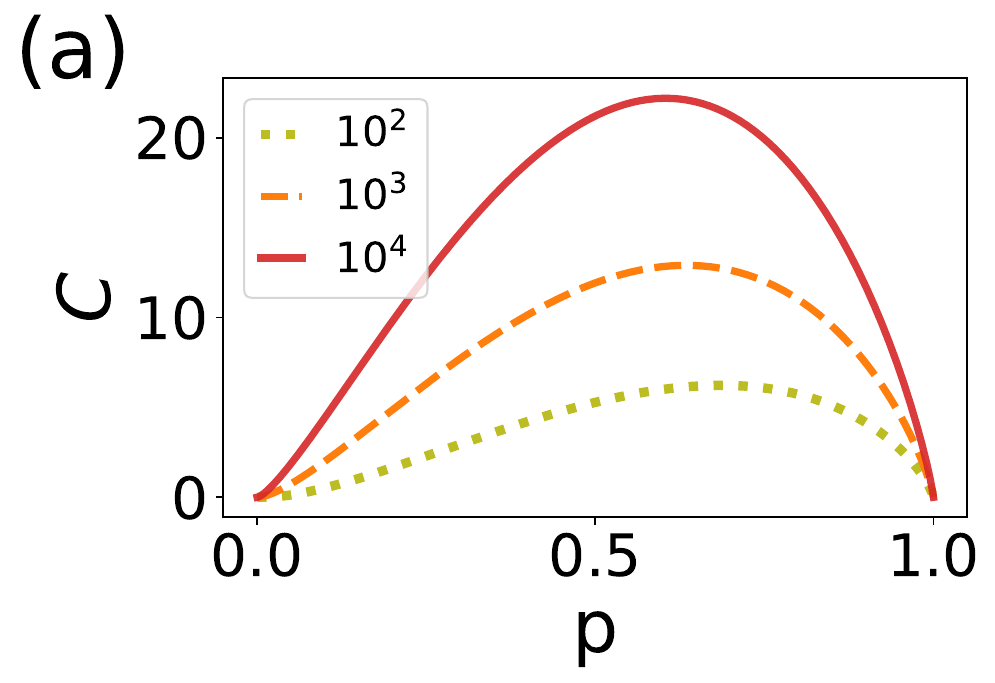}
    \hfill
    \includegraphics[width=0.49\columnwidth]{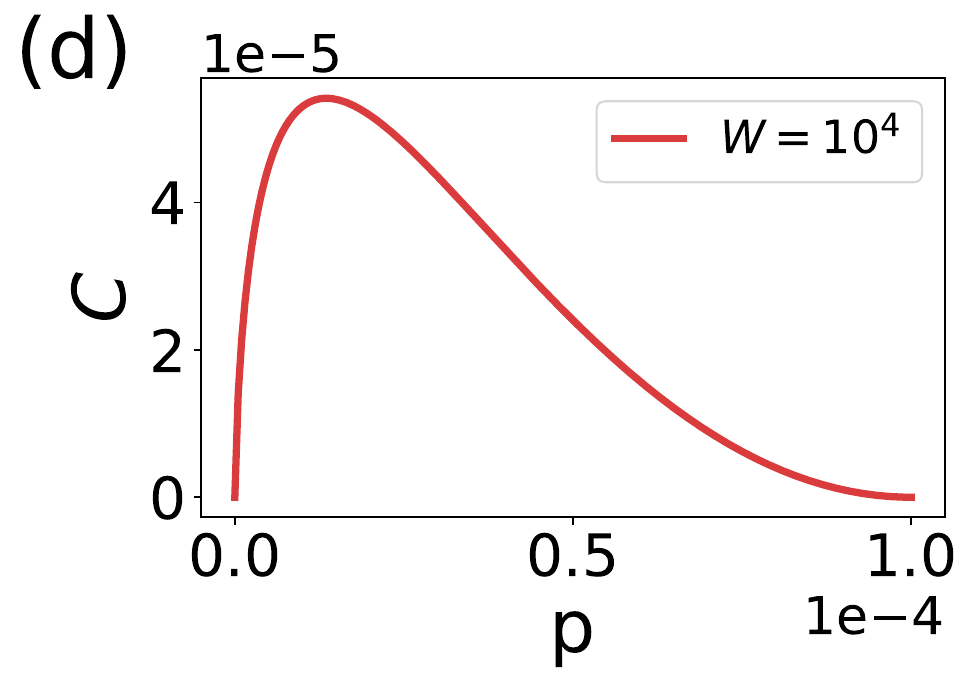}



    \caption{(a) $C(p)$ for a $W$-state system along the path $p_1=p$ and $p_\alpha=(1-p)/(W-1)$ ($\alpha=2,\cdots,W$), for $W=10^2$, $10^3$, and $10^4$. (b) Enlarged view of the small-$p$ region for $W=10^4$.
    The horizontal and vertical axes are scaled by factors of $10^{-4}$ and $10^{-5}$, respectively.}
    \label{fig2}
\end{figure}

To generalize these results, consider a system with $W$ states along the symmetric path $p_1=p$ and $p_\alpha=(1-p)/(W-1)$ for $\alpha \neq 1$. In this case,
\begin{equation}\label{cw1}
C(p)=p(1-p)\left[\ln\frac{p}{1-p}+\ln(W-1)\right]^2 .
\end{equation}
As $W$ increases, the global maximum shifts toward $p=1/2$, while the secondary maximum near $p=0$ becomes negligible [Fig.~\ref{fig2}]. In the asymptotic limit $W \gg 1$, Eq.~\ref{cw1} reduces to $C(p)\simeq p(1-p)(\ln W)^2$, resulting in a maximum at $p=1/2$. The maximum complexity becomes
\begin{equation}\label{cmax}
C_{\max}\simeq \frac14(\ln W)^2 ,
\end{equation}
indicating slow (logarithmic) growth with system size.  Numerical results confirm that Eq.~(\ref{cmax}) remains accurate even for moderately large $W$ (see Table 1).

\section{Case of Boltzmann--Gibbs factor and phase transitions}

A central result is the connection between $C$ and equilibrium thermodynamics. For Boltzmann--Gibbs probabilities $p_{BG}=\exp(-\beta \mathcal{H})/Z$, with $Z=\mathrm{Tr}\,\exp(-\beta \mathcal{H})$ and $\beta=1/(k_B T)$,  $C$ assumes the form
\begin{equation}
C=\beta^2\left(\langle\mathcal{H}^2\rangle-\langle\mathcal{H}\rangle^2\right)
=\beta^2\,\text{Var}(\mathcal{H}).
\end{equation}
Up to a multiplicative constant, $C$ coincides with the heat capacity, being directly related to energy fluctuations. Thus, complexity acquires a thermodynamic interpretation as a measure of information-content fluctuations, implying singular behavior near critical points, including discontinuities at first-order transitions. 

More generally, $C$ is additive for statistically independent systems. For joint probabilities $p_{ij}^{A\cup B}=p_i^A p_j^B$, one has $C^{A\cup B}=C^A+C^B$. Consequently, for $n$ identical independent subsystems, $C^{(n)}=nC^{(1)}$. For Boltzmann--Gibbs systems with short-range interactions, this scaling is consistent with the thermodynamic notion of extensivity.

To illustrate the behavior of $C$ for Boltzmann--Gibbs statistics, we consider the two-dimensional Ising model, which exhibits a continuous phase transition at the critical temperature $T_c$. For the isotropic case $J_x = J_y = 1$, the free energy per spin is given by the Onsager solution~\cite{onsager44} 
\begin{equation}
\label{free}
f(T) = -\frac{1}{8 \pi^2 \beta} \, I(\beta) - \frac{\ln 2}{\beta} \, ,
\end{equation}
where 
\begin{equation}
\label{Ibeta}
\begin{split}
I(\beta) = \int_0^{2\pi} d\theta_1 \int_0^{2\pi} d\theta_2 \,
\ln \big[ \cosh^2(2\beta) - \sinh(2\beta)(\cos\theta_1 + \cos\theta_2) \big] \,. 
\end{split}
\end{equation}
The entropy per particle follows from $s=-(\partial f/\partial T)$, and the critical temperature is $k_B T_c/J=\frac{2}{\ln(1+\sqrt{2})}\approx 2.26919$. The complexity per spin is then obtained from $C=c_{v}/k_B$, where $c_v=T(\partial s/\partial T)$.

\begin{figure}[t]
    \centering

    \includegraphics[width=0.85\columnwidth]{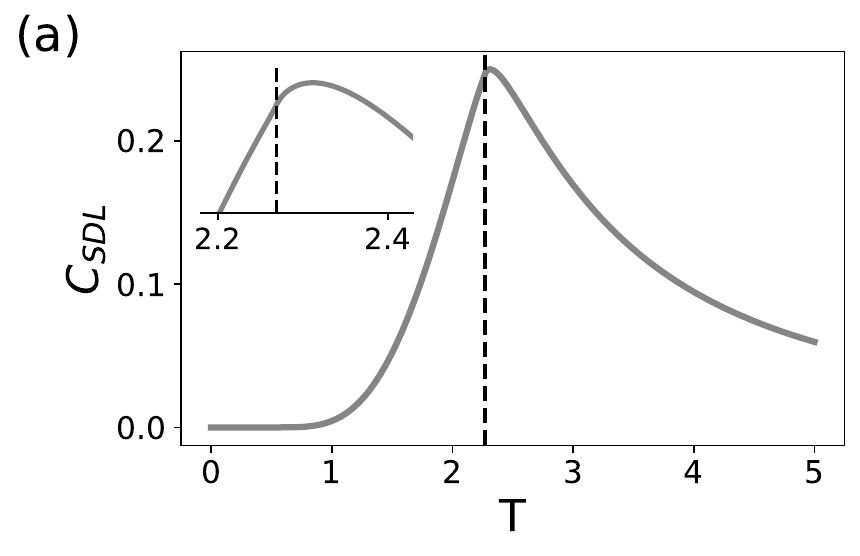}
    
   \includegraphics[width=0.8\columnwidth]{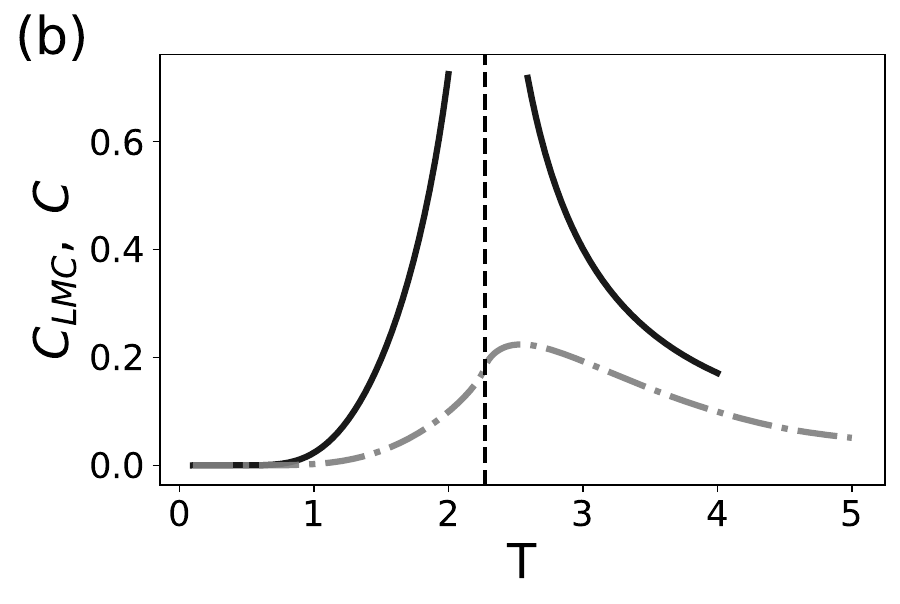}

    \caption{Statistical complexity measures for the two-dimensional Ising model. The dashed vertical line indicates the critical temperature $T_c\approx2.26919$. (a) SDL complexity versus temperature $T$. The inset shows an enlarged view near $T_c$. (b) LMC complexity and $C$ per spin versus $T$. The SDL and LMC measures exhibit maxima near $T_c$, whereas $C$ displays a logarithmic divergence precisely at the critical temperature.}
    \label{fig3}
\end{figure}

Using the entropy per spin $s$, we also compute the SDL and LMC statistical complexities~\cite{landsberg99,calbet95} across the transition (see the Appendix for definitions). For the LMC measure, the disequilibrium per spin can be written as
\begin{equation}
D(T)=\frac{z(T/2)}{z(T)^2}-\frac{1}{2},
\end{equation}
where $z(T)=\exp[-f(T)/k_B T]$ is the partition function per spin. As shown in Fig.~\ref{fig3}, both measures provide good estimates of $T_c$, although their maxima are slightly shifted from the exact critical temperature. By contrast, $C$ reaches its maximum precisely at $T_c$, reflecting the logarithmic divergence of the specific heat in the two-dimensional Ising model.

An upper bound for the MPR statistical complexity~\cite{rosso04} can also be obtained analytically. Since the entropy is maximal at equiprobability, $S\left(\{\frac{p_\alpha+p_e}{2}\}\right)\le S(\{p_e\})$, where $p_e=1/W$, 
the normalized  Jensen--Shannon divergence (see the Appendix) implies
\begin{equation}
C_{\mathrm{MPR}}(T)\le \frac{1}{2}s(T)\,[1-s(T)]\,.
\end{equation}
Therefore, the upper bound of the MPR complexity is proportional to the SDL complexity and qualitatively resembles the LMC measure, so it is not expected to peak exactly at $T_c$.

\section{Chaotic maps}

To assess whether $C$ captures nontrivial dynamical structure beyond equilibrium systems, we analyze the logistic map
\begin{equation}
x_{n+1}=rx_n(1-x_n),
\end{equation}
where $r\in[0,4]$ is the control parameter. For each value of $r$, a time series of size $n=2^{14}$ was generated after discarding transients. The invariant probability distribution of $x_n$ in the interval $[0,1]$ was estimated using 120 equally spaced bins. We compare $C$ with the SDL, LMC, and MPR statistical complexities across chaotic and periodic regimes. 

\begin{figure}[t]
    \centering
    \includegraphics[width=0.49\columnwidth]{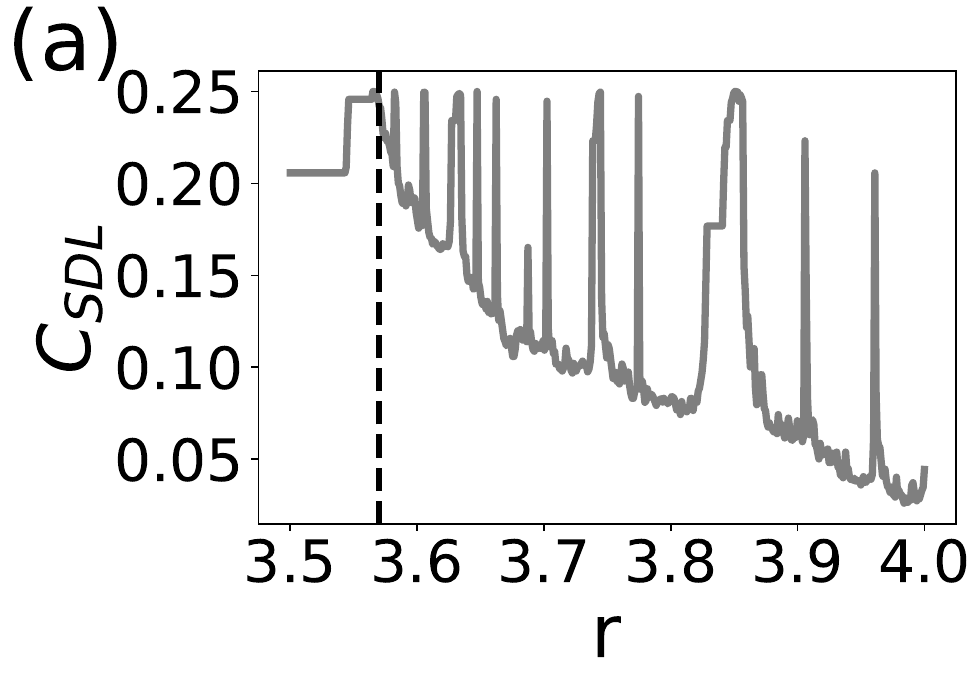}
    \hfill
    \includegraphics[width=0.49\columnwidth]{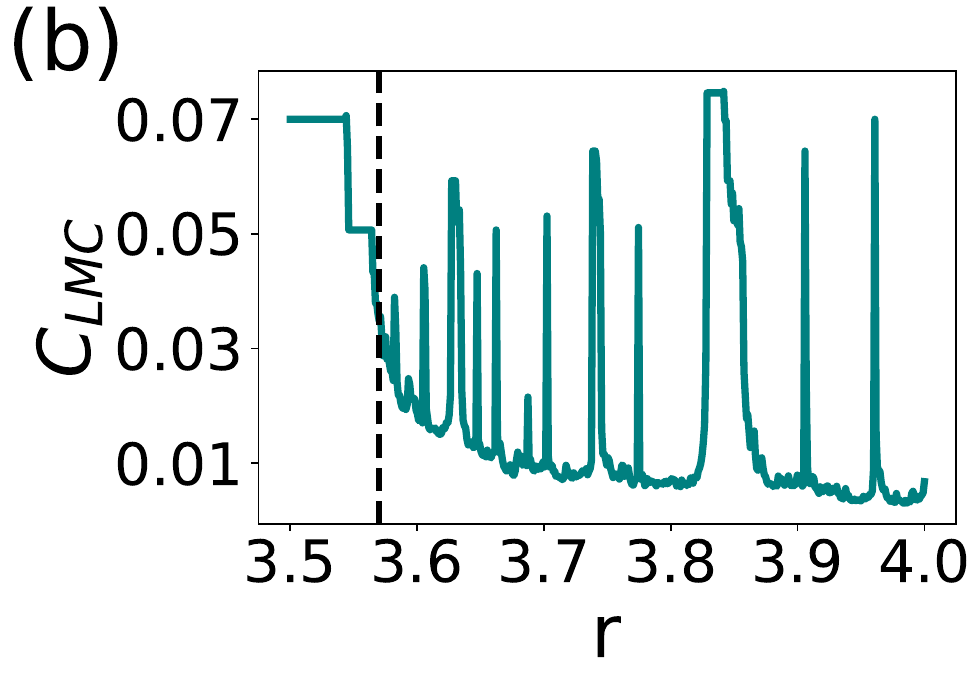}

    \vspace{2mm}

    \includegraphics[width=0.49\columnwidth]{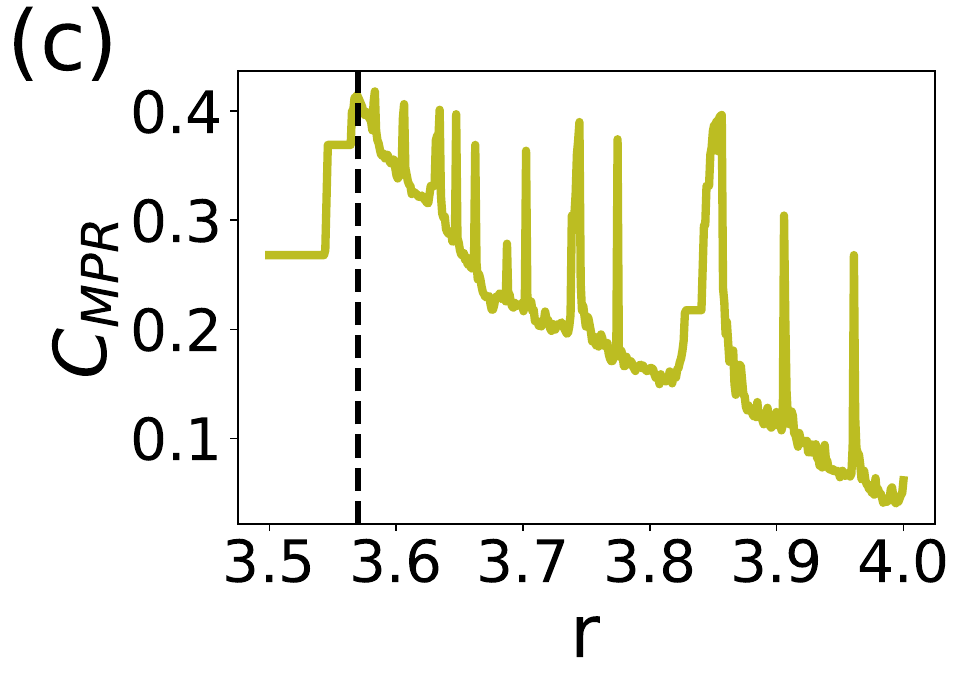}
    \hfill
    \includegraphics[width=0.49\columnwidth]{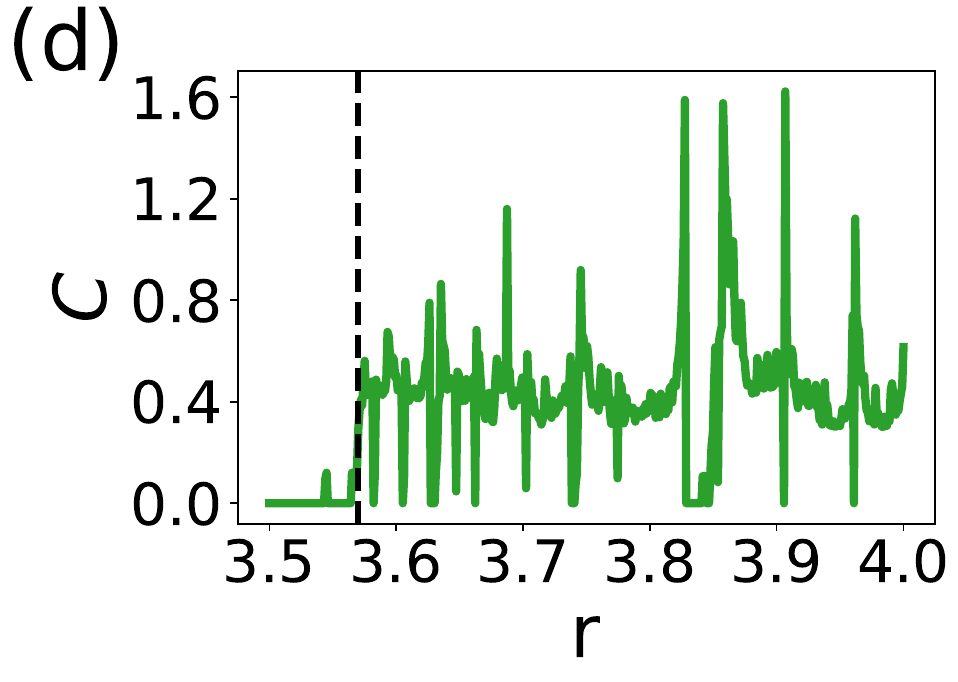}

    \caption{Logistic map complexity. SDL, LMC, MPR, and the present measure $C$ are shown as functions of the control parameter $r$. Probability distributions were estimated from binned time series. The vertical dashed line marks the onset of chaos at $r\approx3.5699$. }
    \label{fig4}
\end{figure}

\begin{figure}[t]
    \centering
    \includegraphics[width=0.49\columnwidth]{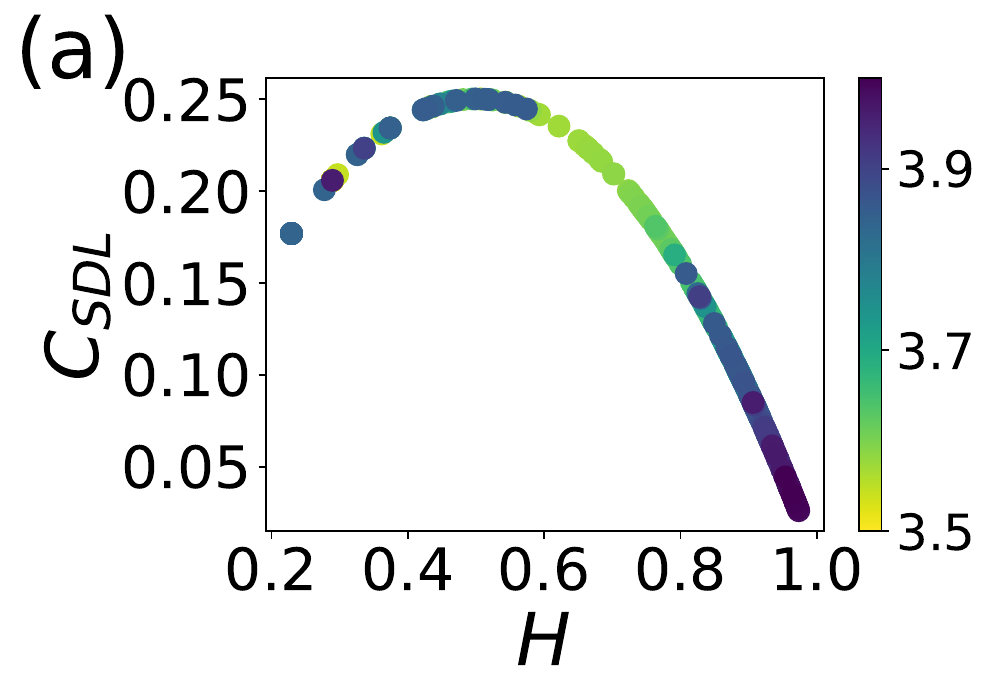}%
    \hfill
    \includegraphics[width=0.49\columnwidth]{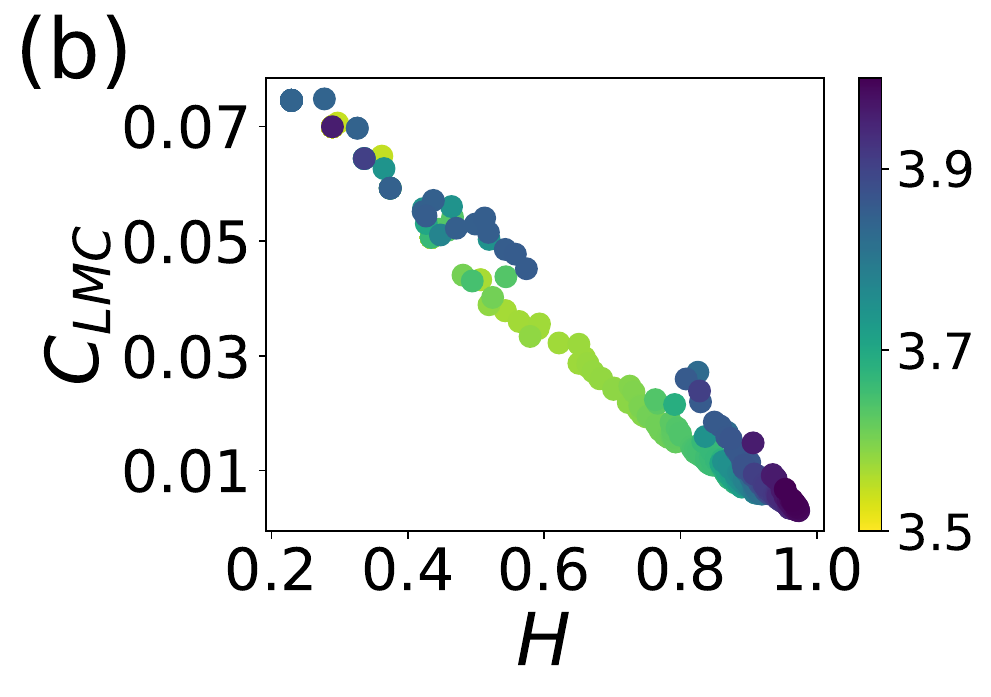}

   \vspace{4mm}

    \includegraphics[width=0.49\columnwidth]{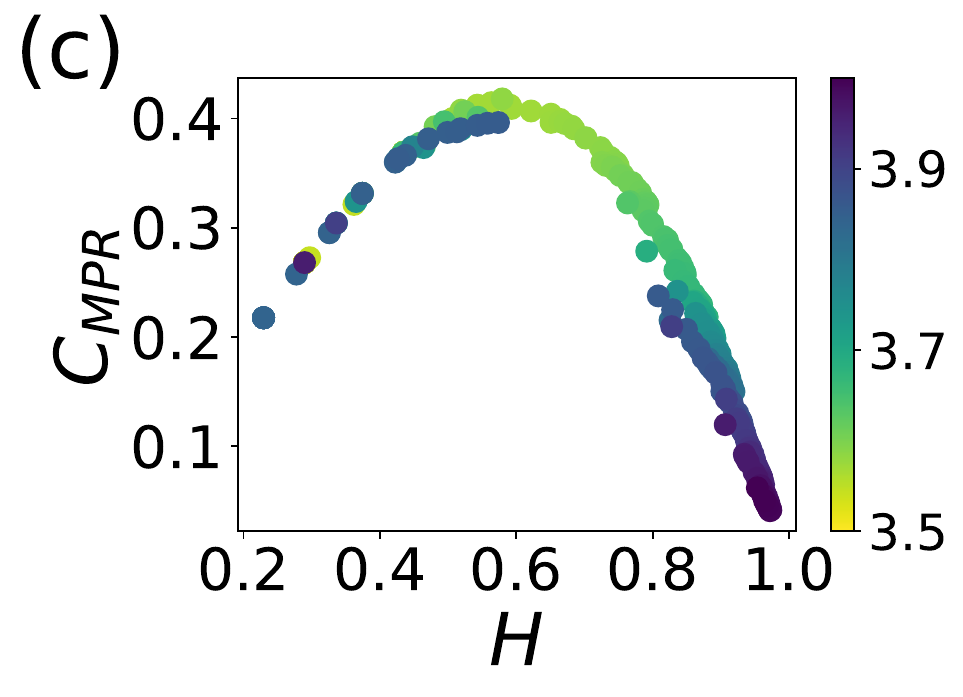}%
    \hfill
    \includegraphics[width=0.49\columnwidth]{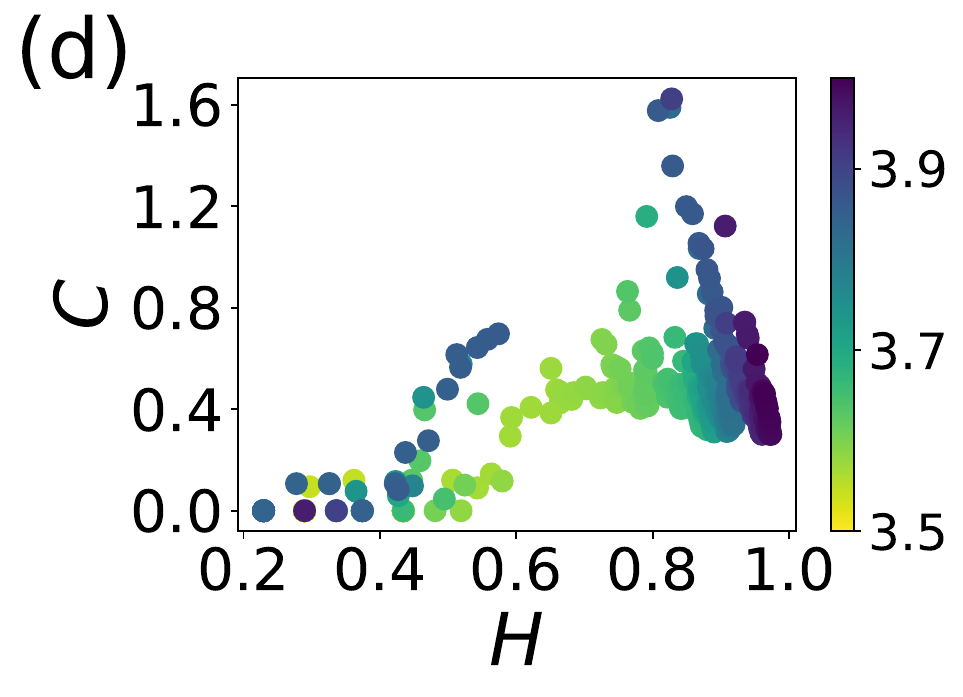}

    \caption{Complexity--entropy plane for the logistic map obtained from binned probability distributions. SDL, LMC, MPR, and the present measure ($C$) are shown as functions of the normalized entropy $H=S/S_{\max}$, with colors indicating the control parameter $r$.}
    \label{fig5}
\end{figure}

Figure~\ref{fig4} shows the SDL, LMC, MPR, and present measure $C$ as functions of $r$. Near $r \approx 3.57$, $C$ increases sharply from values close to zero to a finite plateau, coinciding with the onset of chaos, in agreement with related analyses based on $\sqrt{C}$~\cite{shepard92}. Within the period-three window around $r \approx 3.83$, $C$ exhibits a pronounced minimum, whereas the other measures display maxima. This contrasting behavior indicates that $C$ effectively identifies periodic dynamics and distinguishes them from the surrounding chaotic regime. Compared with the other measures, $C$ also displays a broader dispersion in the $S$-$C$ plane for the logistic map [Fig.~\ref{fig5}]. This broader spread indicates that systems with similar entropy values may exhibit substantially different complexity values according to $C$.

We also analyze the skew tent map,
\begin{equation}
x_{n+1}=
\begin{cases}
x_n/a, & 0 \le x_n < a,\\
(1-x_n)/(1-a), & a \le x_n \le 1,
\end{cases}
\end{equation}
where $a\in(0,1)$ is the control parameter. In contrast to the logistic map, distinct dynamical regimes are not clearly visible in the bifurcation diagram, making the system a natural test case for a permutation-based analysis~\cite{pompe02}. For each value of $a$, a time series of length $n=2^{14}$ was generated after discarding transients, and the permutation entropy was computed using ordinal patterns with embedding dimension $D=5$. 

\begin{figure}[t]
    \centering
    \includegraphics[width=0.48\columnwidth]{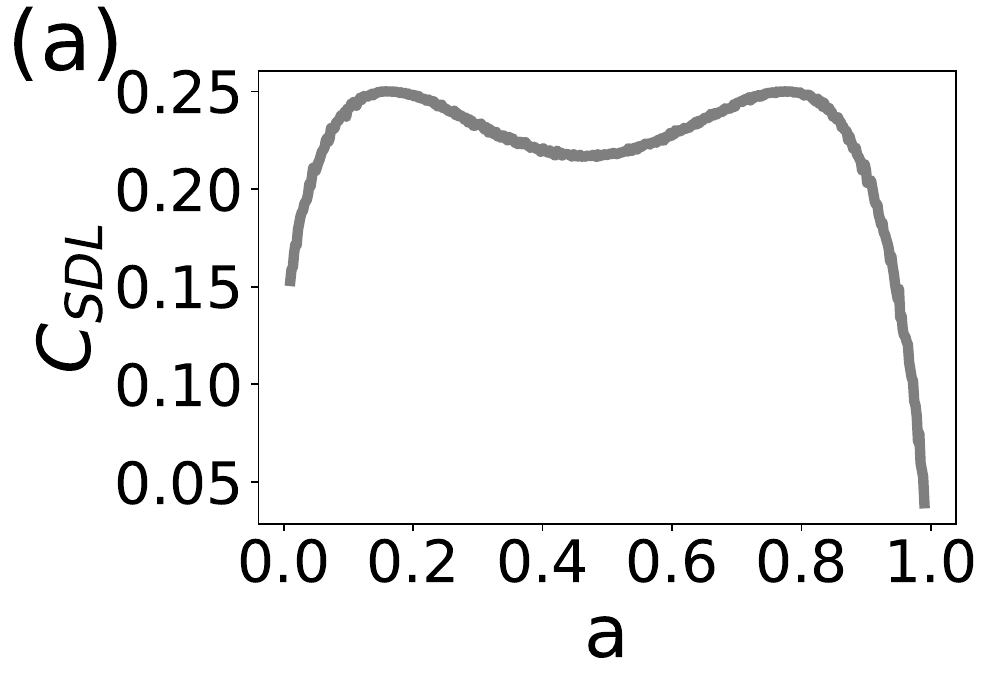}
    \hfill
    \includegraphics[width=0.48\columnwidth]{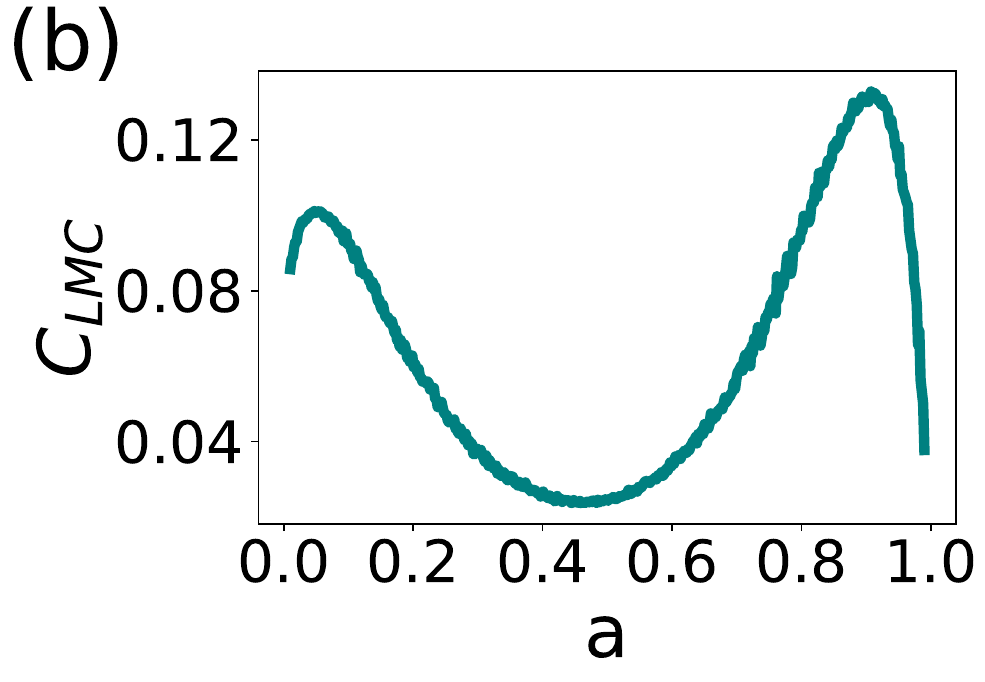}

    \vspace{2mm}

    \includegraphics[width=0.48\columnwidth]{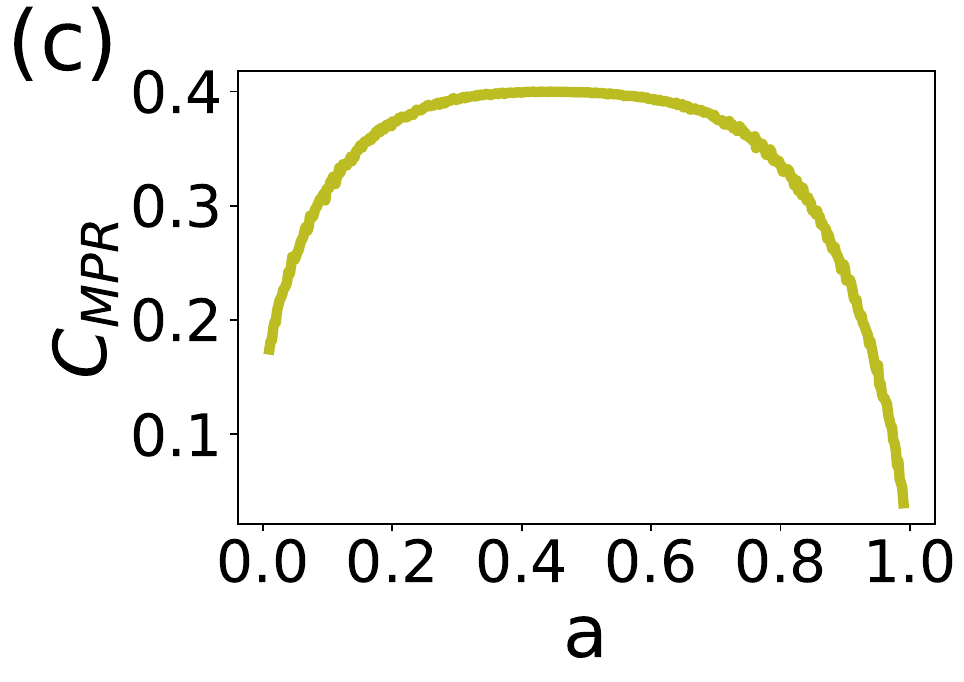}
    \hfill
    \includegraphics[width=0.48\columnwidth]{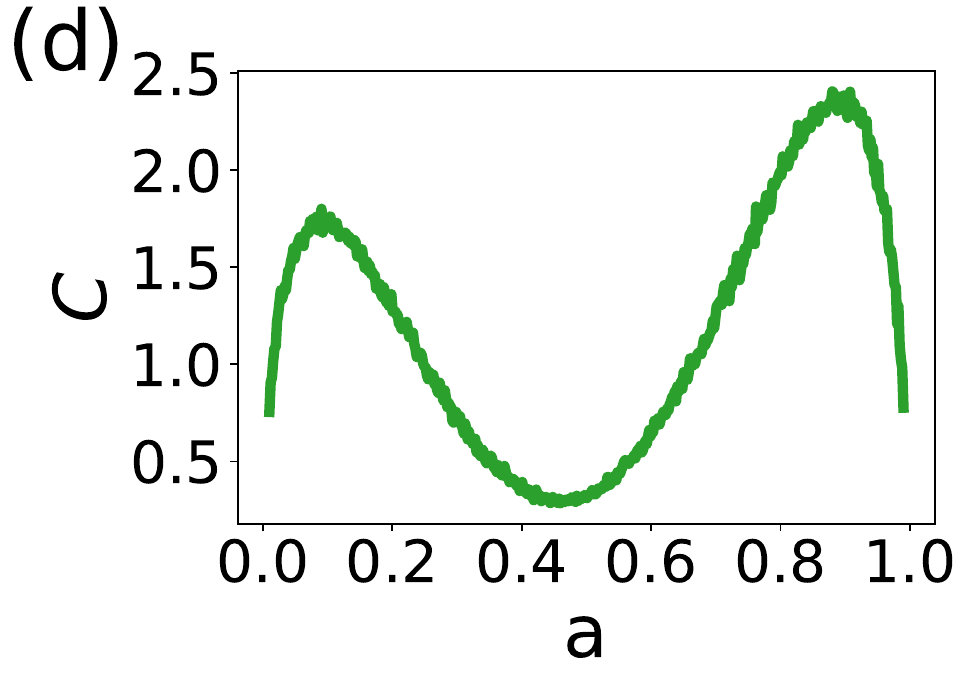}

    \caption{Skew tent map complexity. SDL, LMC, MPR, and the present measure $C$ are shown as functions of the control parameter $a$. Probability distributions were estimated from ordinal patterns with embedding dimension $D=5$ (120 permutations).}
    \label{fig6}
\end{figure}

\begin{figure}[t]
    \centering
    \includegraphics[width=0.48\columnwidth]{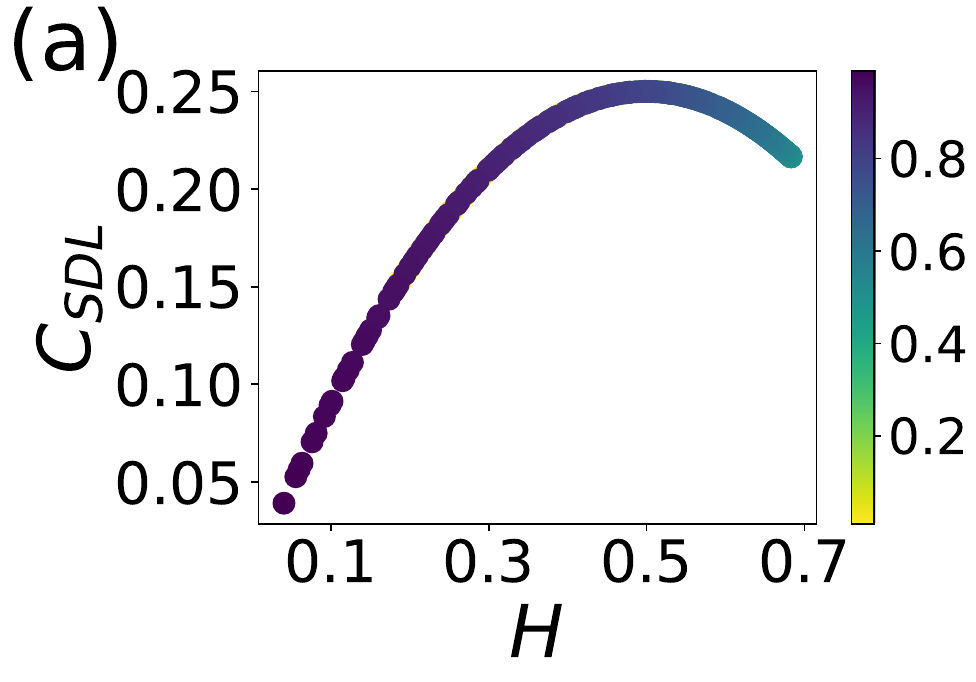}
    \hfill
    \includegraphics[width=0.48\columnwidth]{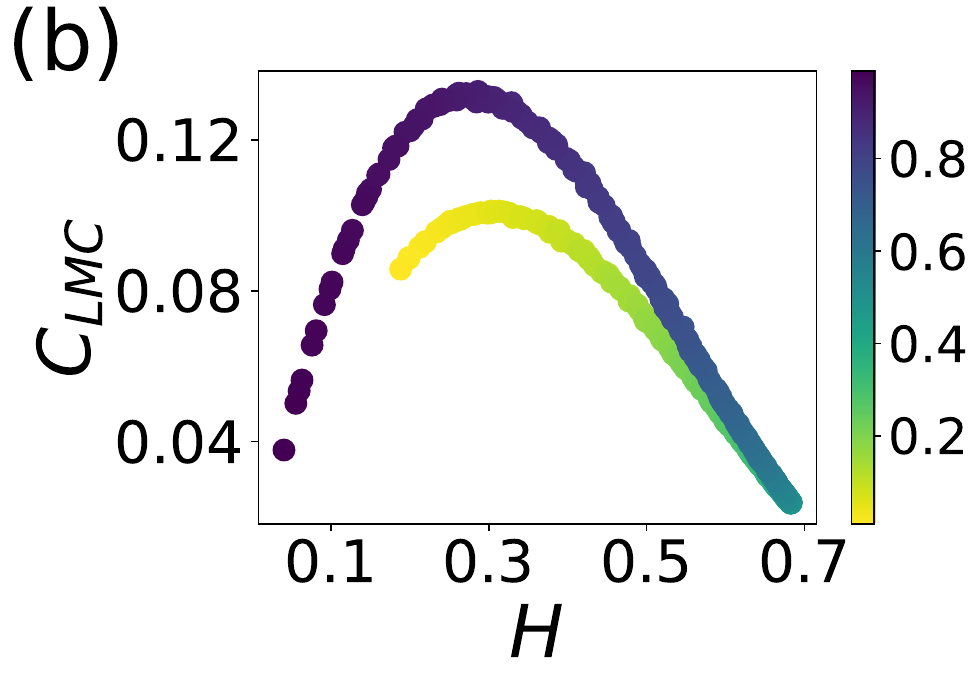}

    \vspace{2mm}

    \includegraphics[width=0.48\columnwidth]{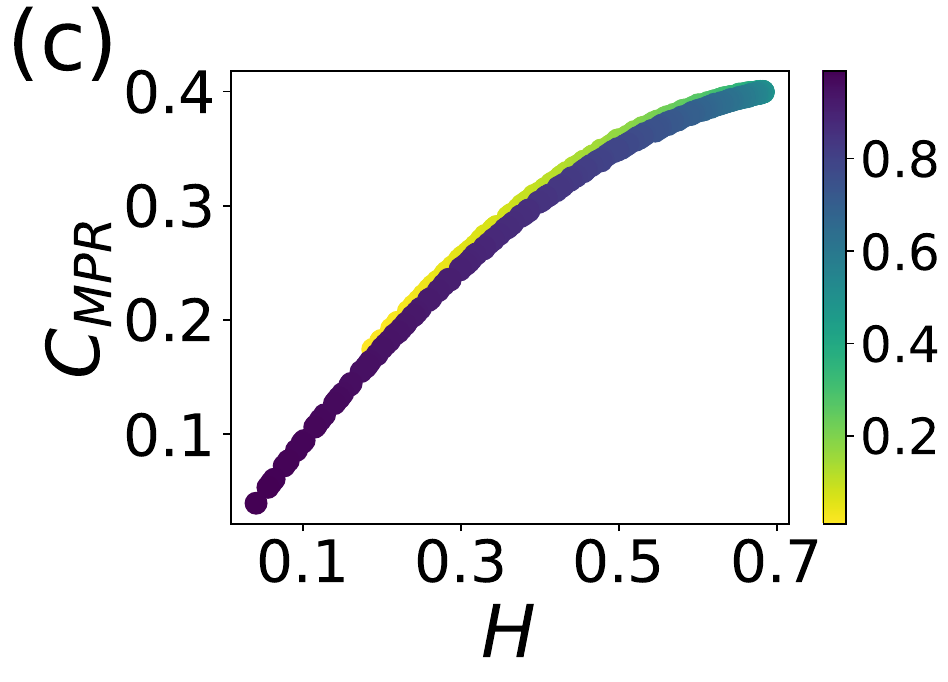}
    \hfill
    \includegraphics[width=0.48\columnwidth]{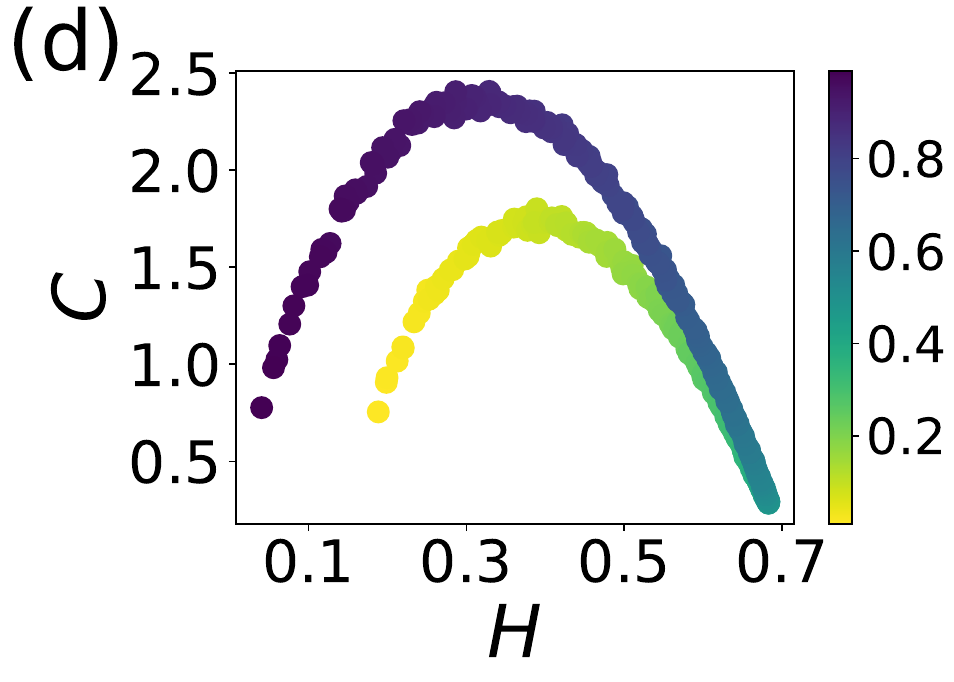}

    \caption{Skew tent map complexity--entropy plane. SDL, LMC, MPR, and the present measure $C$ are shown as functions of the normalized entropy $H=S/S_{\max}$, with colors indicating the control parameter $a$.}
    \label{fig7}
\end{figure}

The corresponding SDL, LMC, MPR, and $C$ complexities are shown in Fig.~\ref{fig6}. LMC and $C$ exhibit similar qualitative behavior, whereas SDL and MPR show a more symmetric structure. Both $C_{\mathrm{LMC}}$ and $C$ exhibit pronounced maxima near the boundaries of the parameter interval, with larger values observed for $a > 0.5$. In the $S$-$C$ plane, the curves for LMC and $C$ split into two branches associated with small and large values of $a$ [Fig.~\ref{fig7}], indicating that systems with similar entropy may nevertheless exhibit distinct dynamical structures. This finding reinforces the idea that entropy alone does not fully characterize the underlying dynamics.

\section{Extensions to other entropic forms and quantum systems}

Because the present formulation is based on fluctuations of informational content rather than on a specific entropy functional, it naturally admits extensions based on generalized entropies, such as Tsallis entropy~\cite{tsallis88,tsallis_book,maadani20}. Such extensions may be relevant mainly for systems with long-range correlations or non-Gaussian statistics. 

Within Tsallis statistics~\cite{tsallis88,tsallis_book}, the $q$-surprise is defined as
\begin{equation}
\ln_q\frac{1}{p_\alpha} = \frac{1-p_\alpha^{\,q-1}}{q-1},
\end{equation}
and the associated Tsallis entropy is
\begin{equation}
S_q[P] = \sum_\alpha p_\alpha
\ln_q\frac{1}{p_\alpha}.
\end{equation}

The generalized complexity is then
\begin{equation} \label{Cq}
C_q = \sum_\alpha p_\alpha \left(\ln_q\frac{1}{p_\alpha} \right)^2 - [S_q[P]]^2,
\end{equation}
which corresponds to the variance of the $q$-surprise~\cite{maadani20}.
$C_q$ reduces to Eq.~\eqref{defC} in the limit $q \to 1$. In the large-$W$ limit, $C_q \sim (1/2)^{2(q-1)}/(q-1)^2$ for $q>1$, whereas for $q<1$ the scaling acquires an additional factor $W^{2(1-q)}$. The parameter $q$
therefore controls the relative weight assigned to low-probability
events.

\begin{figure}[t]
    \centering
    \includegraphics[width=0.48\columnwidth]{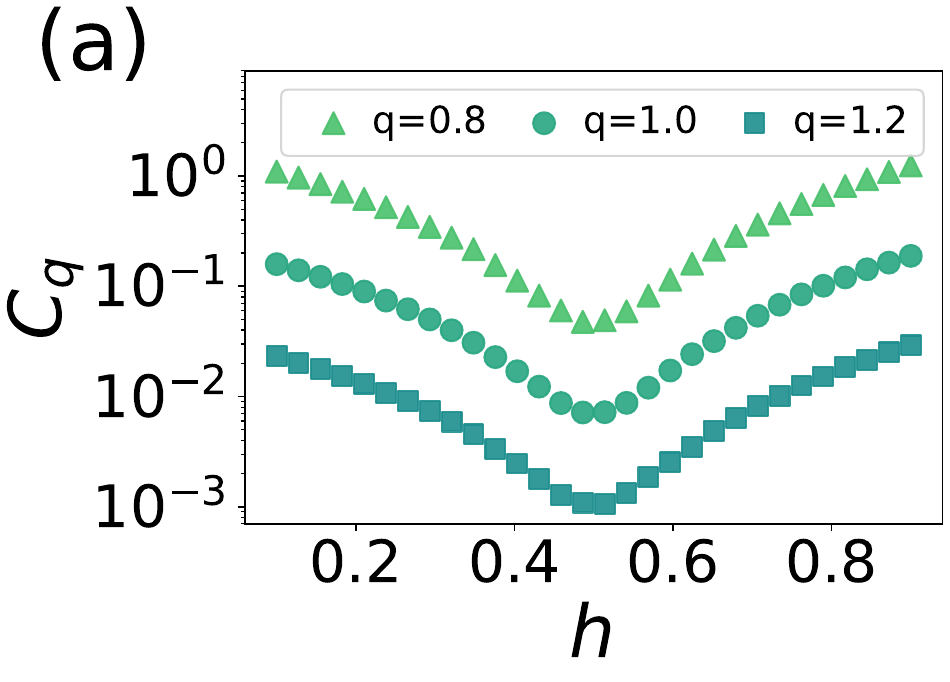}
    \hfill
    \includegraphics[width=0.48\columnwidth]{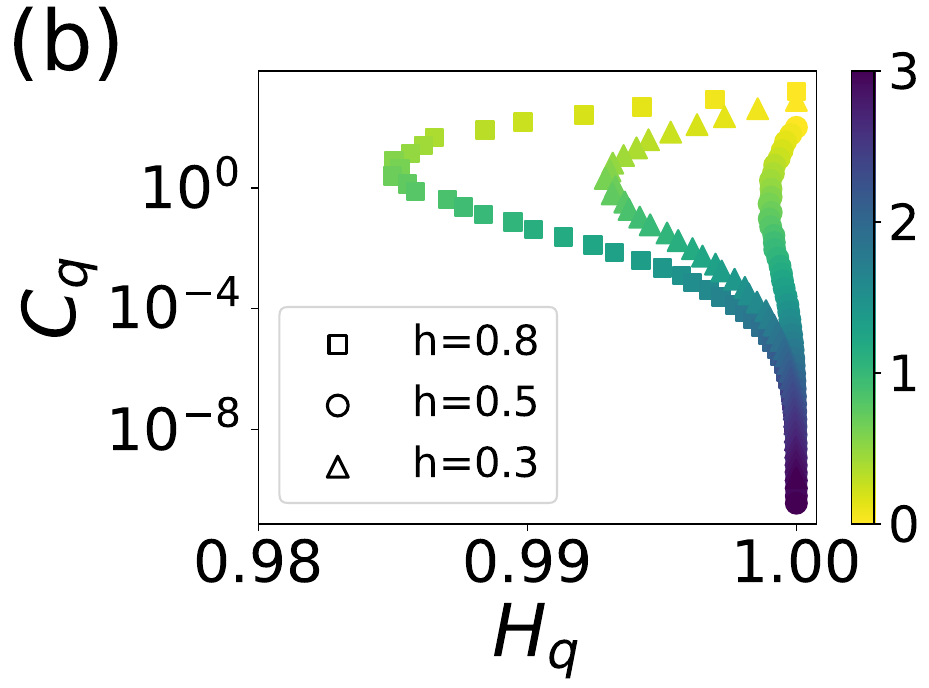}

    \caption{Fractional Gaussian noise complexity. (a) $C_q$ as a function of the Hurst exponent $h$ for selected values of $q$. (b) Complexity--entropy plane showing $C_q$ versus $H_q=S_q/S_q^{max}$ for selected values of $h$. The color scale represents $q$ varying from 0 to 3. Curves correspond to averages over 100 realizations.}
    \label{fig8}
\end{figure}

As a simple illustration of the generalized formulation, we analyze fractional
Gaussian noise (fGn), characterized by the Hurst exponent $h$~\cite{mandelbrot68}. For each value of $h$, we generated time series of length $2^{14}$ and constructed the probability distribution using ordinal patterns with embedding dimension $D=5$ within the permutation entropy framework. As expected, the complexity attains a minimum near $h=1/2$, corresponding to the uncorrelated case, and increases as the dynamics becomes more persistent or anti-persistent [Fig.~\ref{fig8}(a)]. Curves with $q<1$ attain systematically larger values, reflecting the enhanced sensitivity to low-probability ordinal patterns, while $q>1$ suppresses these contributions resulting in smaller complexity values. The generalized complexity may also be analyzed in the associated $S_q$-$C_q$ plane, providing a geometric representation of the interplay between entropy and fluctuations of the $q$-surprise. In this representation, different values of $q$ emphasize different regions of the $S_q$-$C_q$ plane and may improve the visual discrimination between Hurst exponents [Fig.~\ref{fig8}(b)].

The fluctuation-based formulation of $C$ also extends naturally to quantum systems described by a density operator $\rho$ in a Hilbert space. In this case,
\begin{equation}
C=\mathrm{Tr}\left[\rho(\ln\rho)^2\right]
-\left[\mathrm{Tr}(\rho\ln\rho)\right]^2,
\label{quantum}
\end{equation}
corresponding to the variance of the surprisal operator $-\ln \rho$. Its properties and applications will be investigated elsewhere.

\section{Discussion}

Thus, we have provided evidence supporting the variance of the information content as a measure of statistical complexity. Unlike entropy, which characterizes the average surprise, it captures fluctuations around this average. Beyond basic requirements for complexity measures, $C$ exhibits nontrivial maxima near ordered configurations. Such behavior reflects the different sensitivities of ordered and equiprobable configurations to perturbations. Small perturbations of nearly ordered distributions can produce large relative changes in information content, whereas perturbations around the equiprobable state have a much weaker effect. From this perspective, it is plausible that a complexity measure should respond differently in these two regimes.

For Boltzmann–Gibbs systems, $C$ becomes directly related to thermodynamic response functions, providing a natural connection with thermodynamics and phase transitions. In this context, the measure is extensive and attains a maximum at continuous phase transitions---as illustrated for the two-dimensional Ising model. Applications to chaotic maps suggest that $C$ captures nontrivial statistical and dynamical structure, providing complementary insights to those offered by representative statistical complexity measures. The framework admits symbolic representations based on ordinal patterns, where the associated entropy reduces to the permutation entropy~\cite{pompe02,fuentes07}. These features suggest potential applications in the analysis of complex systems.

Despite the encouraging results, the behavior of $C$ in high-dimensional and multiscale systems remains an open question, particularly regarding coarse graining and symbolic representations. Unlike several normalized statistical complexity measures, $C$ is not bounded. In Boltzmann--Gibbs systems, this feature is physically meaningful because it preserves the growth and possible divergence of thermodynamic fluctuations near criticality. Taken together, these results suggest that fluctuations of information content provide a natural and physically grounded basis for statistical complexity and offer a direct connection between information theory, thermodynamics, and complex systems.

\appendix*
\section{Entropy and statistical complexity measures}

All quantities are computed from a discrete probability distribution
$P=\{p_\alpha\}_{\alpha=1}^W$, where $W$ is the number of accessible states.
In the manuscript, $P$ is obtained either from amplitude-based binning or from ordinal patterns (permutation entropy approach). The Shannon entropy is~\cite{shannon48}
\begin{equation}
S[P]=-\sum_{\alpha=1}^W p_\alpha \ln p_\alpha ,
\end{equation}
with normalized form $H[P]=S[P]/\ln W$.
For permutation entropy, the probabilities $p_\alpha$
correspond to the relative frequencies of ordinal patterns of embedding
dimension $d$, with $W=d!$~\cite{pompe02}.

The complexity measure discussed in this work is
\begin{equation}\label{C_apA}
C[P] = \langle(I - S)^2\rangle
      = \sum_\alpha p_\alpha(\ln p_\alpha)^2 - [S[P]]^2,
\end{equation}
which corresponds to the variance of the information content, where $I_\alpha=-\ln p_\alpha$. For Boltzmann--Gibbs statistics, $C[P]$ becomes proportional to the heat capacity, establishing a direct connection with equilibrium thermodynamics. 

For comparison, we also consider other representative statistical
complexity measures. The SDL complexity is~\cite{landsberg99}
\begin{equation}
C_{\mathrm{SDL}}=\Delta[P](1-\Delta[P]),
\end{equation}
where $\Delta[P]=S[P]/S_{\max}$. When the maximum entropy is attained
for the uniform distribution, $\Delta[P]=H[P]$. The LMC complexity is defined as~\cite{calbet95}
\begin{equation}
C_{\mathrm{LMC}}=H[P]D[P],
\end{equation}
where
\begin{equation}
D[P]=\sum_{\alpha=1}^W\left(p_\alpha-\frac{1}{W}\right)^2
\end{equation}
is the disequilibrium function.

The MPR complexity is~\cite{rosso04}
\begin{equation}
C_{\mathrm{MPR}}=H[P]Q_J[P],
\end{equation}
where
\begin{equation}
Q_J[P]=Q_0 J[P,P_e],
\end{equation}
with $P_e$ denoting the uniform distribution and
\begin{equation}
J[P,P_e] = S\!\left[\frac{P+P_e}{2}\right] -\frac{1}{2}S[P] -\frac{1}{2}S[P_e]
\end{equation}
being the Jensen--Shannon divergence. Here, $Q_0$ is a normalization constant chosen such that $0 \le Q_J[P] \le 1$ for all $P$.


\begin{acknowledgments}
The authors thank Marcelo A. Pires (CBPF) for useful suggestions.  
We also thank the Conselho Nacional de Desenvolvimento Cient\'ifico e Tecnol\'ogico (CNPq) and E.M.F.C. thanks the Funda\c{c}\~ao Carlos Chagas Filho de Amparo \`a Ci\^encia e Tecnologia do Estado do Rio de Janeiro (FAPERJ) for the financial support. 
\end{acknowledgments}

\bibliography{refs}

@article{shannon48,
  author    = {Claude E. Shannon},
  title     = {A Mathematical Theory of Communication},
  journal   = {Bell Syst. Tech. J.},
  volume    = {27},
  number    = {3},
  pages     = {379--423},
  year      = {1948},
}

@inproceedings{renyi61,
  author    = {Alfr{\'e}d R{\'e}nyi},
  title     = {On Measures of Entropy and Information},
  booktitle = {Proceedings of the Fourth Berkeley Symposium on Mathematics, Statistics and Probability},
  volume    = {1},
  pages     = {547--561},
  year      = {1961},
  publisher = {University of California Press},
  address   = {Berkeley, California}
}

@article{tsallis88,
  author    = {C. Tsallis},
  title     = {Possible Generalization of {{Boltzmann-Gibbs}} Statistics},
  journal   = {J. Stat. Phys.},
  volume    = {52},
  number    = {1-2},
  pages     = {479--487},
  year      = {1988},
}

@book{tsallis_book,
  author    = {Constantino Tsallis},
  title     = {Introduction to Nonextensive Statistical Mechanics: Approaching a Complex World},
  publisher = {Springer},
  year      = {2009},
  address   = {New York},
  isbn      = {978-0387853581}
}

@article{khinchin53,
  author    = {A. Ya. Khinchin},
  title     = {The Concept of Entropy in the Theory of Probability},
  journal   = {Uspekhi Matematicheskikh Nauk (Russian Mathematical Surveys)},
  volume    = {8},
  number    = {3},
  pages     = {3--20},
  year      = {1953}
}

@Article{shepard92,
	author =       {J. E. Bates and H. K. Shepard},
	title =        {Measuring complexity using information fluctuation},
	journal =      {Phys. Lett. A},
	volume =       {},
	number =       {},
	pages=         {416--425},
	year =		 {1993},   
}

@Article{calbet95,
	author =       {R. L\'opez-Ruiz and H. L. Mancini and X. Calbet},
	title =        {A statistical measure of complexity},
	journal =      {Phys. Lett. A},
	volume =       {209},
	number =       {},
	pages=         {321--326},
	year =		 {1995},   
}

@Article{landsberg99,
	author =       {J. S. Shiner and M. Davison and P. T. Landsberg},
	title =        {Simple measure for complexity},
	journal =      {Phys. Rev. E},
	volume =       {59},
	number =       {2},
	pages=         {1459--1464},
	year =		 {1999},   
}

@Article{rosso04,
	author =       {P. W. Lamberti and M. T. Martin and A. Plastino and O. A. Rosso},
	title =        {Intensive entropic non-triviality measure},
	journal =      {Physica A},
	volume =       {334},
	number =       {},
	pages=         {119--131},
	year =		 {2004},   
}

@Article{ribeiro17,
	author =       {H. V. Ribeiro and M. Jauregui and L. Zunino and E. K. Lenzi},
	title =        {Characterizing time series via complexity-entropy curve},
	journal =      {Phys. Rev. E},
	volume =       {95},
	number =       {062106},
	pages=         {},
	year =		 {2017},  
}

@Article{maadani20,
	author =       {S. Maadani and G. R. M. Borzadaran and A. H. R. Roknabadi},
	title =        {A new generalized varentropy and its properties},
	journal =      {Ural Mathematical Journal},
	volume =       {6},
	number =       {1},
	pages=         {114-129},
	year =		 {2020},  
}

@Article{fuentes07,
	author =       {O. A. Rosso and H. A. Larrondo and M. T. Martin and A. Plastino and M. A. Fuentes},
	title =        {Distinguishing Noise from Chaos},
	journal =      {Phys. Rev. Lett.},
	volume =       {99},
	number =       {},
	pages=         {154102},
	year =		 {2007},   
}

@Article{feldman98,
	author =       {D. P. Feldman and J. P. Crutchfield},
	title =        {Measures of statistical complexity: Why?},
	journal =      {Phys. Lett. A},
	volume =       {238},
	number =       {244252},
	pages=         {},
	year =		 {1998},   
}

@Article{pompe02,
	author =       {C. Bandt and B. Pompe},
	title =        {Permutation Entropy: A Natural Complexity Measure for Time Series},
	journal =      {Phys. Rev. Lett.},
	volume =       {88},
	number =       {17},
	pages=         {},
	year =		 {2002},   
}

@Article{verdu14,
  author  = {I. Kontoyiannis and S. Verd\'u},
  title   = {Optimal lossless data compression: Non-asymptotics and asymptotics},
  journal = {IEEE Trans. Inf. Theory},
  volume  = {60},
  number  = {2},
  pages   = {777--795},
  year    = {2014},
}

@Article{arikan16,
	author =       {E. Arıkan},
	title =        {Varentropy decreases under the polar transform},
	journal =      {IEEE Trans. Inf. Theory},
	volume =       {62},
	number =       {6},
	pages=         {3390-3400},
	year =		 {2016},   
}

@article{onsager44,
  author  = {Onsager, Lars},
  title   = {Crystal Statistics. I. A Two-Dimensional Model with an Order-Disorder Transition},
  journal = {Physical Review},
  volume  = {65},
  number  = {3-4},
  pages   = {117--149},
  year    = {1944},
}

@article{crutchfield89,
  title = {Inferring statistical complexity},
  author = {Crutchfield, James P. and Young, Karl},
  journal = {Physical Review Letters},
  volume = {63},
  issue = {2},
  pages = {105--108},
  numpages = {0},
  year = {1989},
}

@article{anteneodo96,
  title={Some features of the {{LMC}} complexity},
  author={Anteneodo, Celia and Plastino, Angel R},
  journal={Physics Letters A},
  volume={222},
  number={1-2},
  pages={43--49},
  year={1996},
  publisher={Elsevier}
}

@article{ahmed26,
  author       = {Farhan Ahmed and Yuya Jeremy Ong and Chad DeLuca},
  title        = {LogitScope: A framework for analyzing {{LLM}} uncertainty through information metrics},
  journal      = {arXiv preprint arXiv:2603.24929},
  year         = {2026},
  archivePrefix = {arXiv},
  eprint       = {2603.24929},
  note         = {ICLR 2026 Workshop}
}

@article{shternshis24,
  author       = {Andrey Shternshis and Piero Mazzarisi},
  title        = {Variance of entropy for testing time-varying regimes with an application to meme stocks},
  journal      = {Decisions in Economics and Finance},
  volume       = {47},
  pages        = {215--258},
  year         = {2024},
  doi          = {10.1007/s10203-023-00427-9}
}

@article{paolillo21,
  author    = {Antonio Di Crescenzo and Luca Paolillo},
  title     = {Analysis and applications of the residual varentropy of random lifetimes},
  journal   = {Probability in the Engineering and Informational Sciences},
  volume    = {35},
  number    = {4},
  pages     = {680--698},
  year      = {2021},
  doi       = {10.1017/S026996482000025X},
  publisher = {Cambridge University Press}
}

@article{mandelbrot68,
  author  = {B. B. Mandelbrot and J. W. Van Ness},
  title   = {Fractional Brownian motions, fractional noises and applications},
  journal = {SIAM Review},
  volume  = {10},
  number  = {4},
  pages   = {422--437},
  year    = {1968}
}

@article{lloyd01,
  author    = {Lloyd, Seth},
  journal   = {IEEE Control Systems Magazine},
  title     = {Measures of complexity: a nonexhaustive list},
  year      = {2001},
  volume    = {21},
  number    = {4},
  pages     = {7--8}
}

@article{gao21,
  author    = {Gao, Jianbo and Xu, Bo},
  title     = {Complex Systems, Emergence, and Multiscale Analysis: A Tutorial and Brief Survey},
  journal   = {Applied Sciences},
  volume    = {11},
  number    = {12},
  pages     = {5736},
  year      = {2021},
  publisher = {MDPI}
}

\begin{table}[h!]
\centering
\caption{Convergence of the exact maximum of Eq.~(\ref{cw1}) toward the asymptotic expression in Eq.~(\ref{cmax}) as $W$ increases. Here, $p_{\max}$ denotes the value of $p$ at which $C$ is maximized.}
\label{tab1}
\begin{tabular}{@{}rccccccc@{}}
\toprule
$W$            & 2     & 4     & 10    & 100   & 1\,000 & 1\,000\,000 & $10^{12}$ \\
\midrule
$p_{\max}$     & 0.917 & 0.851 & 0.790 & 0.686 & 0.634  & 0.571       & 0.539     \\
$C(p_{\max})$  & 0.439 & 1.023 & 2.058 & 6.227 & 12.900 & 48.710      & 191.867   \\
$C_{\max}$     & 0.120 & 0.480 & 1.325 & 5.302 & 11.929 & 47.717      & 190.878   \\
\bottomrule
\end{tabular}
\end{table}

\end{document}